# DATA ACQUISITION AND UESR INTERFACE OF BEAM INSTRUMENTATION SYSTEM AT SRRC

Jenny Chen, C. J. Wang, C. H. Kuo, K. H. Hu, C. S. Chen, K. T. Hsu, SRRC, Hsinchu, Taiwan

Abstract

Data acquisition systems for the accelerator complex at SRRC composed various hardware and software components. Beam signals are processed by related processing electronics, and connect to control system by various type of interfaces in data acquisition front-end. These front-end include VME crates, personal computers and instruments bus adapters. Fast Ethernet connected all elements together with control consoles. User interface is running on control console. Real-time data capture; display and alaysis are supported on control console. Analysis tools based on Matlab scripts are adopted. Hardware and software implementation of the system will be presented. User interface supports are also described.

## 1 INTRODUCTION

The control system of SRRC is essential to operate the light source efficiently. The console level composed control console and control server. Field level computer composed more than 30 VME crates system and several PCs as special devices. Both level computers system is connected by a dedicated control fast Ethernet. Beam instrumentation system composed various beam monitors and associated processing electronics. The most important beam parameters include beam current monitor and closed orbit, beam profile, etc. Various software tools are integrated with the system and provide an efficient way for operation .

## 2 OUTLINE OF THE CONTROL AND BEAM INSTRUMENTATION SYSTEM

### 2.1 SRRC Control System

The control systems are a two-level hierarchy computer system [1]. Upper layer computers include two process computer and many workstations and PCs. Database management, archive and various application programs are executed on process computer. Main purposes of workstations are used as operation console. Bottom layer computers are distributed VME crate controllers that are in charge of the control and data acquisition for accelerator equipment. Both computer layers are connected by local area network. Software environment can be divided into four logical layers. They are device access layer, network access layer, database layer and applications from bottom to up.

Database plays a role as data exchange center for applications and subsystem of accelerator system. Most of the data updated into database ten times per second.

### 2.2 Beam Instrumentation System

Beam instrumentation system composed various monitors and supporting electronics. The system support precision intensity measurement and lifetime calculation. Orbit measurement, synchrotron radiation monitor and destructive monitor. All devices are controlled by VME based local controller. Control console can access these devices through user interface. Synchrotron radiation monitor is control by a PC and connected to control system via Ethernet. Beam diagnostic instrumentation systems provide necessary electron beam parameters for commissioning, routine operations and beam physics studies.

### 2.3 User Interface

The Common Desktop Environment (CDE) is used for user interface development It is an integrated graphical user interface for SRRC control system, combining X Window System, OSF/Motif, and the CDE technologies. Motif GUI Builder and Code Generator, UIM/X GUI builder is a used to generated various user interface. It enables software developers to interactively create, modify, test and generate code for the user interface portion of their applications. To satisfied various requirements, some application are development in LabVIEW based environment. For fast prototyping, user can customize user's application in Malta environment. Control system supported various database access MEX files. User's can access the database directly in Matlab. For IEEE-488 based interface, fast Ethernet based GPIB adapter was support. And GPIB/2 interface is also included to allow user access these instruments within Matlab.

## 3 SPECIFIC APPLICATIONS

### 3.1 BPM System

The BPM system consists of 57 BPMs equip with switched electrode processing electronics. The data acquisition is done by a dedicated VME crate equip with 128 channels 16 bit ADC channels. The VME host send raw data to orbit feedback system via reflective memory. Averaged data are updated to control database every 10 times per second with resolution around one micron.

### 3.2 Orbit Display Utilities

To provide better orbit observation, a Motif based orbit display utility was developed. This utility provides basic orbit display and difference orbit display. Several display attributes can be selected by users, such as full scale range, persistent display mode enable/disable, average number, save, print, … etc. Update rate of the display is up to 10 Hz. This is very useful for routine operation and various machine studies.

### 3.3 Turn-by-Turn BPM System

To support various beam physics studies, several BPMs equip with log-ratio processor for turn-by-turn beam position measurement. Multi-channel VME form factor digitizer with 12-bit ADC acquires turn-by-turn beam position. A server program running on VME crate manages the data acquisition of beam position. Client program running on control console and Motif based GUI are used for user interface.

### 3.4 SmartLink System to Acquire Beamline Data

To acquire data come form remote site, a SmartLink based system was setup. The link is used a private Ethernet as field bus. The Ethernet are connecting to a PMC Ethernet module installed on VME host. The SmartLink data acquisition module is used to acquire data from beamline monitor with high resolution, include photon flux (Io) monitor and photon BPM blades current. Update rate is about 2 time per second with 20-bit resolution. This slow update rate is the major disadvantage.

### 3.5 Synchrotron Radiation Monitor Interface

The synchrotron radiation monitor is used to measure beam profile. It consists an optics and high resolution CCD. To acquired profile information, a PC acquires the image from CCD camera, analysis the profile and extract profile parameters. The local display is also broadcast via facility-wide machine status CATV system. Server program running on PC serve the data request form control console. A client program running on control console can access information of this PC by the help of LabVIEW program or Matlab MEX files running on control console.

### 3.6 Gap Voltage Modulation Study Support

RF gap voltage modulation was adopted to relieve the effect of longitudinal coupled-bunch instability in routinely operation of the storage ring at SRRC. Systematic measurements were done recently to investigate the mechanism why RF gap voltage modulation can do this. The experimental setup is shown as Figure 1. Gap voltage modulation frequency is about 2fs (50 kHz). The rolled off frequency of the LLRF gap voltage regulation loop is about 7 kHz. A function generator in VME form factor generates the modulation sinusoidal wave. This generator integrates with control system to satisfy the requirement of routine operation. Frequency and amplitude can be adjusted on control console. Modulation signal is injected after loop filter and added with correction signal in the RF gap voltage regulation loop. HP4396A spectrum/network analyzer observes beam spectrum form BPM sum signal.

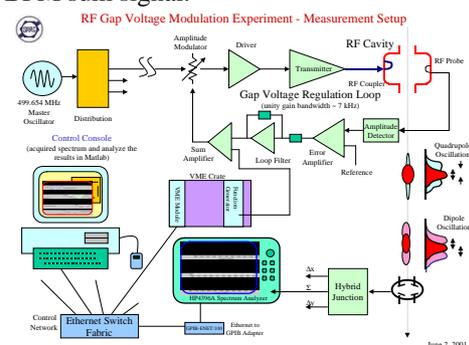

Figure 1. Experimental setup for RF gap voltage modulation study.

National Instrument GPIB/ENET-100 controller connect the spectrum analyzer to control network. Control console support database accesses MEX-file that allows Matlab to read and set the modulation parameters. GPIB MEX-file allows Matlab to read and write the GPIB devices via GPIB/ENET-100 controller. The experimental data can be acquired directly into Matlab. The gap voltage modulation is effective due to the modulation frequency is far beyond the unity gain cutoff frequency of the gap voltage regulation loop. Modulation amplitude is about 10 percent of total gap voltage in routine operation. Experiments sequence is programmed by a simple Matlab script running on control console that select frequency scan range as well as modulation amplitude. The spectrum analyzer can select to measure upper or lower sideband either 1fs or 2 fs synchrotron oscillation frequency. Figure 2 shown the measured result and clear show that the instability is suppressed near 50 kHz.

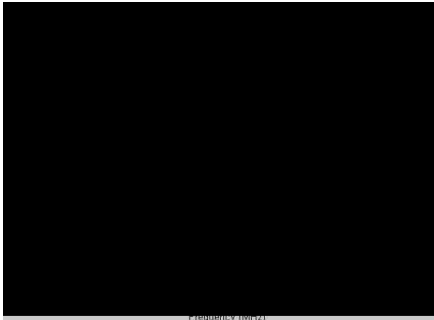

Figure 2. Typical scan spectrum of RF gap voltage modulation study.

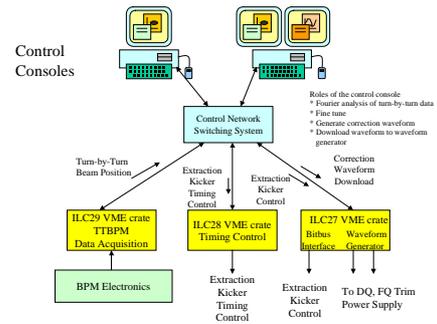

Figure 3. Tune acquisition and correction system.

### 3.7 Booster Tune Measurement and Correction System

The electron beam is accelerated form 50 MeV to 1.5 GeV at the booster synchrotron with 50 msec ramping time. Tune is an important index indicated the tracking performance of the White circuit based power supply system. Tune information was obtained by tune measurement system; it provides the $\nu x$ and $\nu y$ during energy ramping. Tune variation in the energy ramping are correlated to the tracking performance of three families White circuit. Optimized lattice can be obtained by the help of measured tune for the booster synchrotron to get a better working point for efficient operation. An application program was developed to automatic measure and to correct tune variation. The stored beam are excited by extraction kicker to perform damped betatron oscillation, trigger timing and field strength of the kicker is set properly as function of beam energy to ensure sufficient beam excitation and without kill the stored beam. Beam motion signal are picked up by stripline and process by log-ratio BPM electronics [2]. A transient digitizer at VME crate records betatron oscillation of the stored beam. A sever programs running on VME crate coordinate the operations of data acquisition. Client program running on control consoles is invoked by a Matlab script file to perform data acquisition and analysis. Data analysis including Fourier analysis, peak identification and visualization. The tune correction signal is generated and downloads to waveform generator located on VME crate. Figure 3 shown the system structure of the tune measurement and correction system [3]. The Matlab script is used to generate correction waveform by the measured tune and measured sensitivity matrix between tune and quadrupole setting. Figure 4 shown that the tune during ramping with and without correction. Tune variation during ramping can be reduced drastically by this feed-forward correction.

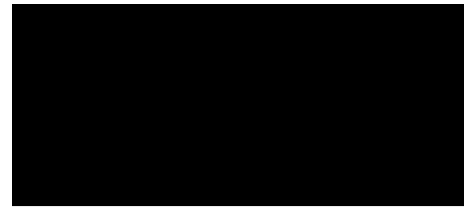

(a)

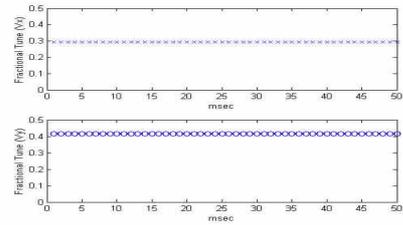

(b)

Figure 4. Tune variation during energy ramping; (a) before correction, (b) after correction.

## 4 SUMMARY

Data acquisition and user interface of SRRC are summary in this report. These systems are essential for the operation of accelerator system for machine tuning, various feedback, machine study, etc. The system is evoluted continually with technology advanced.